\documentclass[10pt]{aastex}

\usepackage{amsmath,natbib}
\usepackage{rotating}
\usepackage{emulateapj5}
\usepackage{graphicx}

\def\spose#1{\hbox to 0pt{#1\hss}}
\def\lta{\mathrel{\spose{\lower 3pt\hbox{$\mathchar"218$}}
     \raise 2.0pt\hbox{$\mathchar"13C$}}}
\def\gta{\mathrel{\spose{\lower 3pt\hbox{$\mathchar"218$}}
     \raise 2.0pt\hbox{$\mathchar"13E$}}}

\newcommand{\ha}{\text{H}$\alpha$}
\newcommand{\hii}{\text{H}{\sc ii}} 

\shorttitle{An upper limit on anomalous dust emission at 31~GHz in the diffuse cloud [LPH96]201.663+1.643}
\shortauthors{C.~Dickinson et al.}

\begin{document}

\title{An upper limit on anomalous dust emission at 31~GHz in the diffuse cloud [LPH96]201.663+1.643}

\author{C.~Dickinson\altaffilmark{1}, S.~Casassus\altaffilmark{2},
J.~L.~Pineda\altaffilmark{2,3}, T.~J.~Pearson\altaffilmark{1},
A.~C.~S.~Readhead\altaffilmark{1} and R.~D.~Davies\altaffilmark{4}}
\altaffiltext{1}{Dept. of Astronomy, California Institute of
Technology, M/S 105-24, Pasadena, CA 91125, U.S.A.}
\altaffiltext{2}{Departamento de Astronom{\'\i}a, Universidad de
Chile,  Casilla 36-D, Santiago, Chile.}
\altaffiltext{3}{Argelander-Institut f\"ur Astronomie, Universit\"at
Bonn, Auf dem H\"ugel 71, D-53121, Bonn, Germany.}
\altaffiltext{4}{Jodrell Bank Observatory, University of Manchester,
Lower Withington, Macclesfield, Cheshire, SK11 9DL, U.K.}

\begin{abstract}
[LPH96]201.663+1.643, a diffuse H{\sc ii} region, has been reported to
be a candidate for emission from rapidly spinning dust grains. Here we
present Cosmic Background Imager (CBI) observations at $26-36$~GHz
that show no evidence for significant anomalous emission. The spectral
index within the CBI band, and between CBI and Effelsberg data at
1.4/2.7~GHz, is consistent with optically thin free-free emission. The
best-fitting temperature spectral index from 2.7 to 31~GHz,
$\beta=-2.06 \pm 0.03$, is close to the theoretical value,
$\beta=-2.12$ for $T_{e}=9100$~K. We place an upper limit of
$24\%~(2\sigma)$ for excess emission at 31~GHz as seen in a $6\arcmin$
FWHM beam. Current spinning dust models are not a good fit to the
spectrum of LPH96. No polarized emission is detected in the CBI data
with an upper limit of $2\%$ on the polarization fraction.
\end{abstract}




\section{Introduction}

To extract precise cosmological information from high-sensitivity
cosmic microwave background (CMB) data, the Galactic foregrounds must
be accurately known over a wide frequency range. Galactic radiation,
which comprises 3 well-known components (synchrotron, free-free and
vibrational dust), is the most important on angular scales $\gtrsim
1\arcdeg$. However, many authors have detected an additional
``anomalous'' component in the frequency range $\sim 10-60$~GHz whose
origin is still not understood. The anomalous emission, first detected
in the COBE-DMR maps at 31- and 53-GHz, was initially thought to be
due to free-free emission owing to its correlation with FIR maps and
its spectral index \citep{Kogut96a,Kogut96b}. It soon became clear
that free-free emission could not account for this ``excess'' due to
the lack of \ha~emission. Observations of the NCP region at 14.5- and
32-GHz \citep{Leitch97} clearly show the anomalous emission and the
tight correlation with $100~\mu$m maps. Since then, anomalous emission
has been detected by a number of authors \citep{deOliveira-Costa02,
deOliveira-Costa04, Banday03, Lagache03, Finkbeiner04a, Finkbeiner04b,
Casassus04, Watson05, Davies06} but still little is known about the
physical mechanism that produces it. Candidates include spinning dust
grains \citep{Draine98a,Draine98b}, magnetic dust emission
\citep{Draine99}, flat-spectrum synchrotron \citep{Bennett03b} and
free-free emission from very hot electrons \citep{Leitch97}.

The first targeted search for spinning dust emission from individual
objects was carried out by \cite{Finkbeiner02}. Using the Green Bank
140ft telescope, with a resolution of $6 \arcmin$, they made scans of
10 dust clouds to look for the spectral signature of spinning dust
grains i.e. a sharp rise from low frequencies up to a peak at $\sim
20$~GHz. They saw a rise in the flux density over the frequency range
$5-10$ GHz for 2 objects: [LPH96]201.663+1.643 (henceforth LPH96) and
LDN1622, both strongly correlated with FIR maps. The spectrum of
LDN1622 appears to fit a spinning dust model remarkably well, a result
that has recently been confirmed with CBI data at 31~GHz
\citep{Casassus06}. LPH96 is the brighter of the two clouds and is
classed as a diffuse \hii~region \citep{Lockman96}.

In \S\ref{sec:cbi} we present CBI total intensity and polarization
observations of LPH96 in the range $26-36$~GHz. The mapping capability
of CBI provides an angular resolution $\sim 6\arcmin$ (similar to
Green Bank observations). In \S\ref{sec:radio_maps} we compare low
frequency radio maps at 1.4 and 2.7~GHz with CBI data at 31~GHz, both
in the image and Fourier planes. \S\ref{sec:gb_fir} discusses the
Green Bank data of \cite{Finkbeiner02} and the correlations with the
$100~\mu$m dust template. Discussion and conclusions are given in
\S\ref{sec:discussion}.


\section{CBI data}
\label{sec:cbi}

The Cosmic Background Imager (CBI) is a 13-element interferometer
situated on the high altitude Chajnantor site in Chile. In a compact
configuration, baseline lengths range from 1.0m to 4.0m corresponding
to angular scales between $\sim 6\arcmin$ and $\sim 30\arcmin$. The
CBI covers the frequency range $26-36$~GHz in 10 1~GHz
channels. Observations of LPH96 (RA(J2000)=06:36:40,
DEC(J2000)=+10:46:28)  were taken on the nights of 15,17 and 19
November 2002 with a combined integration time of $\sim 3000$~s. Each
receiver measures either left (L) or right (R) circular polarization,
thus each baseline measures either total intensity (LL or RR) or
polarization (LR or RL), which are combined in the $u,v$-plane to give
Stokes $I,Q,U$. We assume that circular polarization, $V=0$. A longer
integration of $\sim 3600$~s was made in total intensity mode (all
receivers measuring L only) on 14 January 2003.

The data were reduced using similar routines to those used for CMB
data \citep{Pearson03,Readhead04a,Readhead04b}. Each 8-min integration
on source was accompanied by a trail field, separated by 8 mins in RA,
observed at the same hour angle for subtraction of
ground-spillover. The overall absolute calibration scale is tied to a
Jupiter temperature of $T_{\rm J}=(147.3 \pm 1.8)$~K
\citep{Readhead04a}. Secondary calibrators (Tau-A, Jupiter, 3C274)
were used to estimate a further uncertainty of $\sim 2\%$ in the gains
on a given night. We therefore assign a total calibration uncertainty
of $3\%$.

\begin{figure*}[!h]
\centering \includegraphics[width=0.4\textwidth, angle=0]{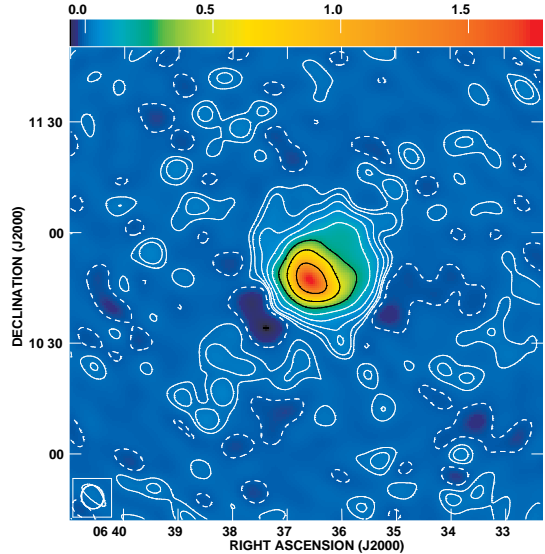}
\caption{CBI 31~GHz total intensity CLEANed map of LPH96. The uniform weighted synthesized beam FWHM is $6\arcmin.5 \times 5\arcmin.9$. The primary beam  ($45\arcmin.2$ FWHM) has not been corrected for in this image. Contours are at $-0.5 (dashed),0.5,1,2,4,8,16,32,64\%$ of the peak flux density, 1.79 Jy/beam.} 
\label{fig:cbi_map}
\end{figure*}

The final CBI CLEANed total intensity map is shown in
Fig.~\ref{fig:cbi_map}. The excellent $u,v$ coverage provides robust
mapping of both compact and extended emission on scales up to $\sim
30\arcmin$ within the primary beam of $45\arcmin.2$ FWHM at
31~GHz. The peak flux density at 31~GHz is $1.79$~Jy/beam centered on
LPH96 but with some extended emission, mainly to the NW with a
deconvolved angular size of $\sim 20\arcmin$. The noise level is
$\approx 10$~mJy/beam. Low level extended emission is also detected
outside the FWHM of the primary beam, particularly to the SE of
LPH96. Two elliptical Gaussians can account for the majority of the
flux in the CLEANed map with an integrated flux density (after
correcting for the primary beam) of $7.57 \pm 0.4$~Jy and residual rms
of $\approx 30$~mJy. The 2 components are 4.1~Jy centered on the peak
of LPH96 with a deconvolved angular size of $9\arcmin.7 \times
6\arcmin.3$ and an extended component with angular size $20\arcmin.2
\times 17\arcmin.0$ containing 3.5~Jy. By splitting the bands into low
and high frequencies, maps at 28.5 and 33.5~GHz were
produced. However, due to the different $u,v$ coverage, the maps have
different resolution and spatial scales making it difficult to compare
the flux densities of extended sources. Nonetheless, the peak flux
density in a restored $6\arcmin$ FWHM beam was 1.75 and 1.71 Jy/beam
at 28.5 and 33.5~GHz, respectively. This implies a flux spectral
index\footnote{The temperature spectral index $\beta$ is related to
the flux density spectral index $\alpha$ by $\beta = \alpha -2$ in the
Rayleigh-Jeans limit.} $\alpha =-0.14\pm 0.19$ over this range.

The polarization mapping capability of the CBI has been demonstrated
from deep CMB observations
\citep{Cartwright05,Readhead04b,Sievers05}. The Stokes $Q$ and $U$
images of LPH96 have a r.m.s. noise level of $\sim 14$~mJy/beam, with
a synthesized beam of $7\arcmin.9 \times 6\arcmin.5$ (FWHM). No
significant polarization is detected above the noise from this region
on these angular scales. From the noise-corrected polarization
intensity map we place a $3 \sigma$ upper limit on the polarization
intensity of 34~mJy. This corresponds to a polarization fraction upper
limit of $2\%$ of the peak. For the extended emission to the NW, with
a brightness of $\sim 0.2$~Jy/beam, the upper limit increases to
$\approx 10\%$.

\section{Comparison with low frequency radio maps}
\label{sec:radio_maps}

At lower frequencies, the all-sky surveys at 408~MHz (Haslam et
al. 1982), 1420~MHz (Reich \& Reich 1986) and 2326~MHz (Jonas et
al. 1998) do not have sufficient angular resolutions ($51\arcmin$,
$35\arcmin$ and $20\arcmin$ FWHM, respectively) to allow a reliable
comparison with CBI data. Instead we use data from the Effelsberg
100-m telescope at 1408~MHz \citep{Reich97} and 2695~MHz
\citep{Fuerst90}. The data\footnote{The Effelsberg survey data were
downloaded from the MPIfR sampler survey website:
http://www.mpifr-bonn.mpg.de/survey.html} are fully-sampled background
subtracted continuum maps with beams of FWHM $9\arcmin.4$ and
$4\arcmin.3$ respectively. The 1.4~GHz map has a peak brightness
temperature $T_{b}=6.88$~K and LPH96 is almost unresolved
(Fig.~\ref{fig:multi_maps}a). The 2.7~GHz map
(Fig.~\ref{fig:multi_maps}b) has a peak $T_{b}=3.38$~K and shows the
extension to the NW detected in the CBI image. After smoothing to a
common resolution of $9\arcmin.4$, the peak brightness at 2.7~GHz is
$T_{b}=1.86$~K, which corresponds to a temperature spectral index
$\beta_{1.4}^{2.7}=-2.02 \pm 0.11$ (we assume a calibration
uncertainty of 5\% in the Effelsberg data).

To allow a comparison with CBI data, the 2.7~GHz map is ``observed''
with the CBI $u,v$ coverage. This involves sampling the fourier
transform of the 2.7~GHz map after multiplying the image by the CBI
primary beam (and also making a correction for the smoothing due to
the Effelsberg beam). The CLEANed image of the 2.7~GHz simulated map
is shown as contours in Fig.~\ref{fig:multi_maps}b. The image is a
good match to the CBI map with a peak flux density of $2.06 \pm
0.10$~Jy/beam and an integrated flux density for the LPH96 region of
$7.6 \pm 0.4$~Jy. The peak flux densities at 2.7 and 31~GHz in the CBI
beam correspond to a flux density spectral index $\alpha=-0.06 \pm
0.03$.

We also compare data in the Fourier plane by simulating the CBI
observation given an input map (as in the image analysis), but
comparing the visibilities directly. This allows direct comparison of
CBI data with multi-frequency maps without the problems of
deconvolution and incomplete $u,v$ coverage. The noise in each
visibility can be treated as independent so we can simply compute the
slope of the best-fitting linear relationship between the visibilities
at the two frequencies. The Pearson's correlation coefficient,
$P=0.93$, indicates a strong correlation between CBI data and the
$2.7$~GHz map. The fitted slopes from 31~GHz to 1.4 and 2.7~GHz are
$2.06\pm 0.12$~mK/K and $7.13 \pm 0.42$~mK/K respectively. This
corresponds to spectral indices of $\beta_{1.4}^{31}=-2.00\pm 0.04$
and $\beta_{2.7}^{31}=-2.02\pm 0.04$. The slopes for each CBI channel
can be calculated in the same way, but are less useful since each
frequency samples different regions of the $u,v$-plane. Nevertheless,
for correlations with the 2.7~GHz data, we get $\beta=-2.19 \pm 0.14$
over the range $26-36$~GHz, in agreement with the image analysis.


\section{Comparison with Green Bank data and FIR maps}
\label{sec:gb_fir}

The Green Bank data presented by \cite{Finkbeiner02} consist of
$48\arcmin$ long scans at 5,8.25 and 9.75 GHz, smoothed to a
resolution $6\arcmin$ FWHM. The scans were chopped at $12\arcmin$ and
a smooth baseline subtracted to remove ground (sidelobe) and
atmospheric contamination. This makes it difficult to obtain accurate
flux densities for extended structures.

However, \cite{Finkbeiner02} do find a tight correlation with the
\cite{Schlegel98} $100~\mu$m dust map which can be used to estimate
the flux density. If the emissivity is constant within a given region,
this allows comparison of data with different resolutions and/or
observing strategies. The temperature-corrected $100~\mu$m dust map is
shown in Fig.~\ref{fig:multi_maps}c with a peak brightness of
479~MJy/sr at $6\arcmin.1$ resolution. The averaged emissivities from
\cite{Finkbeiner02}, referenced to the $100~\mu$m map, are $2770.5 \pm
13.8$, $1361.0 \pm 48.0$ and $1262.5 \pm 49.5~\mu$K/(MJy/sr) at 5,
8,25 and 9.75~GHz, respectively. The Pearson's correlation was
$P=0.91$ indicating the overall similarity between the radio and
$100~\mu$m data. The dust map was scaled with these values and
observed with the CBI beam. For a restored CLEAN beam of $6\arcmin$
(FWHM), the peak flux densities were $1.97 \pm 0.23$, $2.64 \pm 0.10$,
$3.38 \pm 0.14$~Jy at 5, 8.25 and 9.75~GHz, respectively.

To test whether there is a significant variation in
$100~\mu$m-referenced emissivity between extended and compact emission
(which would invalidate the comparison), the correlation coefficient
was calculated within a restricted $u,v$-range corresponding to either
small or large angular scales. The correlation coefficient at 31~GHz
was $41.6 \pm 1.2~\mu$K/(MJy/sr) and did not change significantly
($\lesssim 5\%$) for angular scales between $6\arcmin$ and
$30\arcmin$. Joint cross-correlations with multiple maps were not
attempted since the structure in the radio and FIR maps is similar and
would not allow a reliable separation of components.


\section{Discussion and conclusions}
\label{sec:discussion}

\begin{figure*}[!h]
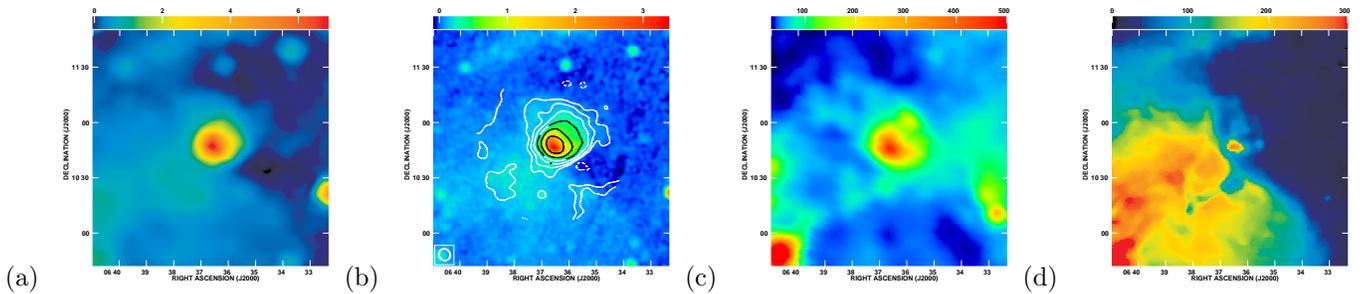

\centering
(a)~ \includegraphics[width=0.2\textwidth, angle=0]{f2a.eps}
(b)~ \includegraphics[width=0.2\textwidth, angle=0]{f2b.eps}
(c)~ \includegraphics[width=0.2\textwidth, angle=0]{f2c.eps}
(d)~ \includegraphics[width=0.2\textwidth, angle=0]{f2d.eps}
\caption{Multi-frequency maps of the LPH96 region, covering the same angular scale as Fig.~\ref{fig:cbi_map}, at their original resolutions (see text). (a) Effelsberg 21cm continuum map (units of K). (b) Effelsberg 2.7~GHz continuum map (units of K) overlaid with a CBI simulation of this data. The simulated map has been CLEANed and primary beam corrected down to $10\%$ of the peak. Contours are at $-2(dashed),2,4,8,16,32,64\%$ of the peak intensity. (c) \cite{Schlegel98} $100~\mu$m temperature-corrected dust map (units of MJy/sr). (d) SHASSA continuum-subtracted \ha~map at $4\arcmin$ resolution (units of Rayleigh).}
\label{fig:multi_maps}
\end{figure*}

To quantify the contribution of dust emission at GHz frequencies, the
free-free emission must be accurately known. The analysis of
\cite{Finkbeiner02} relied on \ha\ data to place limits on the amount
of free-free radiation, which was found to be dominant at
5~GHz. However, the effects of dust extinction can clearly be seen in
the \ha\ map \citep{Gaustad01} shown in
Fig.~\ref{fig:multi_maps}d. LPH96 is visible at a level of $\sim
250$~R. For optically thin emission at 2.7~GHz, and assuming a typical
electron temperature of $T_{e}\sim 7000-10000$~K, the \ha-to-free-free
conversion is $\approx 1~$mK/R \citep{Dickinson03}. The measured
electron temperature from radio recombination lines (RRL) is
$T_{e}=9100$~K \citep{Shaver83}. This would predict 0.25~K of
free-free emission based on \ha\ intensities and would require an
absorption factor of $> 10$ to match the radio flux density. Making
predictions based on \ha\ are therefore unreliable. Observations of
RRLs provide a complementary and clean method of separating the
free-free thermal emission, without the effects of dust
absorption. \cite{Lockman96} measure a H126$\alpha$ line temperature
for LPH96 of $T_{L}=(24 \pm 2.6)$~mK at $9\arcmin$
resolution. Assuming optically thin emission in local thermodynamic
equilibrium, for $T_{e}=9100$~K, this would predict $4.9\pm 0.5$~K of
free-free emission at 1.4~GHz in a $9\arcmin$ beam. This implies that
the radiation at 2.7~GHz remains optically thin down to at least
1.4~GHz, thus allowing reliable extrapolation from 2.7~GHz to higher
frequencies. There is low-level extended emission to the SE of LPH96
that has a similar spectrum suggesting that the emission in this
entire region is dominated by optically thin free-free emission.  The
WMAP data\footnote{WMAP all-sky maps available from
http://lambda.gsfc.nasa.gov/.}  \citep{Bennett03a}, smoothed to
$1\arcdeg$ resolution, have a spectral index $\alpha=-0.22 \pm 0.14$
over the range $20-40$~GHz. A detailed comparison of WMAP data and CBI
data is not attempted due to the mismatch of resolutions.

\begin{figure*}[!h]
\centering
\includegraphics[width=0.50\textwidth, angle=0]{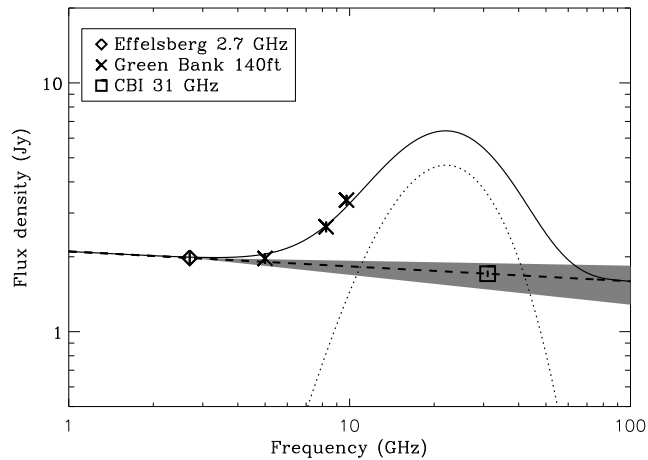}
\caption{Spectrum of LPH96. Data points are in terms of the peak flux density in a $6\arcmin$ FWHM Gaussian beam, after being observed with the CBI beam. The Effelsberg 2.7~GHz point ({\it diamond}) is assumed to be due to free-free emission only. The dashed line is the emission extrapolated from 2.7~GHz with a flux density spectral index $\alpha=-2.06$. The shaded area represents the possible range for the free-free extrapolation from 2.7~GHz with $-0.12< \alpha <-0.02$ (see text). The CBI flux density at 31 GHz ({\it square}) is seen to be dominated by free-free emission alone. The Green Bank 140 ft data, at 5,8.25 and 9.75~GHz ({\it crosses}), are evaluated using correlations with the $100~\mu$m dust template (see text). A spinning dust model for the Warm Neutral Medium \citep{Draine98b}, scaled to fit the Green Bank data, is shown as the {\it dotted line}. {\it Solid line} is the sum of free-free and spinning dust components.} 
\label{fig:spectrum}
\end{figure*}

The spectrum of LPH96, in terms of the flux density within a
$6\arcmin$ (FWHM) Gaussian beam centered on LPH96, is shown in
Fig.~\ref{fig:spectrum}; $1.98 \pm 0.10$~Jy at 2.7~GHz and $1.71 \pm
0.05$~Jy at 31~GHz. The Green Bank points were estimated by
calculating the flux density in the $6\arcmin$ CBI beam of the
$100~\mu$m map scaled by the factors as measured by
\cite{Finkbeiner02} (see \S\ref{sec:gb_fir}). The gray shaded area
shows the allowed range for the free-free model based on extrapolating
from 2.7~GHz using $-0.02< \alpha <-0.12$. The range is based on the
flattest derived spectral index ($\alpha=-0.02$) and the theoretical
value ($\alpha=-0.12$); the best-fitting value from the image
analysis, $\alpha=-0.06 \pm 0.03$, is in the middle of this range. The
5~GHz point is close to the free-free model, but at $8-10$~GHz, the
Green Bank data show a significant excess that cannot be reconciled
with the 2.7 and 31~GHz data points. A spinning dust model for the
Warm Neutral Medium \citep{Draine98b}, scaled in amplitude to fit the
Green Bank data, is depicted in Fig.~\ref{fig:spectrum}. From this,
one would expect to see considerable emission in the $26-36$~GHz CBI
band, well above the free-free emission, which is not seen. It is
interesting to note that the Galactic plane survey of
\cite{Langston02} at 8.35~GHz ($11\arcmin.167$ beam) detected LPH96
with a peak flux density of 3.69~Jy and integrated flux density of
6.74~Jy. But no detection was made at 14.35~GHz ($8\arcmin$ beam),
with a detection limit of 2.5~Jy. Given the spectral rise observed
from $5-10$~GHz, one would expect $\approx 5$~Jy at 14.35~GHz.

If the spectrum is indeed increasing from 5 to 10 GHz, then the lack
of anomalous emission at CBI frequencies suggests that either i)
current spinning dust models are not a good fit for this particular
cloud or ii) the emission is due to another
mechanism. \cite{McCullough02} have suggested that the rising spectrum
could occur if there was a dense optically thick ultracompact H{\sc
ii} region within the cloud, which would exhibit a rising spectrum
($\alpha \approx 2$) with a flux density in the range $1.6-4.4$~Jy at
15~GHz. High resolution data are required to definitively rule out
this model, but it is unlikely given the CBI flux
density.\footnote{D. Finkbeiner et al. (priv. comm) have recently
re-observed LPH96 using the 100m Green Bank Telescope (GBT) in the
frequency range $5-18$~GHz. They could not reproduce the rising
spectrum as seen with the 140ft telescope.}

From the image analysis, we can constrain the non-free-free
contribution at 31~GHz by subtracting the free-free model based on the
extrapolation of 2.7~GHz data. A spectral index $\alpha=0.06\pm 0.03$
accounts for essentially all the emission seen at 31~GHz and is close
to the canonical value of $\alpha=-0.1$ for free-free emission at GHz
frequencies. Adopting the theoretical value, $\alpha=-0.12$ for
$T_{e}=9100$~K, the predicted flux density at 31~GHz in a $6\arcmin$
beam becomes $1.48 \pm 0.07$ Jy. This leaves $0.23 \pm 0.09$~Jy, or
$14\%$ of the total 31~GHz flux density, that could be due to an
additional anomalous component. At this level the dust emissivity
would be $\approx 6~\mu$K/(MJy/sr), within the range observed at high
latitudes (e.g. \cite{Banday03}). It is therefore possible that there
is a non-negligible anomalous dust component in this H{\sc ii} region
that is not detected because of the much brighter thermal
emission. Given the sensitivity to the assumed spectral index, we do
not claim a significant detection of anomalous emission at 31~GHz. We
place an upper limit 0.41~Jy ($2\sigma$) on the anomalous emission at
31~GHz in a $6\arcmin$ beam, which corresponds to $24\%$ of the total
flux density.

The lack of polarization seen in the CBI data are consistent with
free-free emission. If $14\%$ of the emission at 31~GHz was indeed
anomalous, then the polarization of this component is $\sim 10\%$
($2\sigma$). This largely rules out emission from aligned grains of
strongly magnetic material that is expected to be highly polarized
\citep{Draine99}.

More data are required to clarify the origin(s) of the anomalous
emission, both at high and low Galactic latitudes, and to investigate
the conditions in which it occurs. For LPH96, data in the range
$5-20$~GHz are required to investigate further the anomalous emission
reported by \cite{Finkbeiner02}. A detailed understanding of
anomalous/spinning dust emission will be important for modeling and
removal of CMB foregrounds at frequencies $\lesssim 100$~GHz,
particularly if they exhibit significant polarization.


\acknowledgements{We gratefully acknowledge support from the Kavli
  Operating Institute and thank B. Rawn and S. Rawn Jr for their
  continuing support. We are also grateful for the support of M. and
  R. Linde, C. and S. Drinkward and the provost, president, and PMA
  division chairman of the California Institute of Technology. The CBI
  was supported by NSF grants 9802989, 0098734 and 0206416. We
  acknowledge the use of the Legacy Archive for Microwave Background
  Data Analysis (LAMBDA). Support for LAMBDA is provided by the NASA
  Office of Space Science. We acknowledge the use of NASA's SkyView
  facility (http://skyview.gsfc.nasa.gov) located at NASA Goddard
  Space Flight Center. The Southern H-Alpha Sky Survey Atlas (SHASSA)
  is supported by the NSF. CD thanks Patricia Reich for maintaining
  the MPIfR image retrieval facility and making the Effelsberg data
  available. We thank Doug Finkbeiner for informing us of new GBT
  observations of LPH96. CD thanks Barbara and Stanley Rawn Jr for
  funding a fellowship at Caltech. SC acknowledges support from
  FONDECYT grant 1030805, and from the Chilean Center for Astrophysics
  FONDAP 15010003. JLP acknowledges the grant MECESUP UCH0118 given by
  the Chilean Ministry of Education.}

\bibliographystyle{apj}
\bibliography{refs.bib}

\end{document}